\documentclass[aps,prb,twocolumn]{revtex4-1}

\usepackage{bm} % bold math

\begin{document}

\title{Comment on ``Weyl fermions and the anomalous Hall
effect in metallic ferromagnets''}

\author{David Vanderbilt}
\affiliation{Department of Physics and Astronomy, Rutgers
University, Piscataway, New Jersey 08854-8019, USA}

\author{Ivo Souza}
\affiliation{Centro de F\'{\i}sica de Materiales and DIPC,
Universidad del Pa\'{\i}s Vasco, 20018 San Sebasti\'an,
and Ikerbasque Foundation, 48011 Bilbao, Spain}

\author{F.D.M. Haldane}
\affiliation{Department of Physics, Princeton University, Princeton,
NJ 08544, USA}

\date{\today}

\begin{abstract}
We point out that, contrary to an assertion by Chen, Bergman and Burkov
[Phys.\ Rev.\ B {\bf 88}, 125110 (2013)],
the non-quantized part of the intrinsic anomalous
Hall conductivity can indeed be expressed as a
Fermi-surface property even when Weyl points are present in the
bandstructure.
\end{abstract}

\maketitle

\newcommand{\beq}{\begin{equation}}
\newcommand{\eeq}{\end{equation}}
\newcommand{\kk}{{\bf k}}

%%=========================================================================
\marginparwidth 2.7in
\marginparsep 0.5in
\def\dvm#1{\marginpar{\small DV: #1}}
\def\ism#1{\marginpar{\small IS: #1}}
\def\dhm#1{\marginpar{\small FDMH: #1}}
\def\scr{\scriptsize}
%%=========================================================================

In a recent paper, Chen, Bergman and Burkov (CBB)\cite{chen-prb13}
challenged the claim that the non-quantized part of the intrinsic
anomalous Hall conductivity (AHC) can be regarded as a Fermi-surface
property.\cite{haldane-prl04} In this Comment, we point out
that CBB misrepresented the previous work, and that the formal
analysis of Ref.~\onlinecite{haldane-prl04}, as well as
subsequent first-principles calculations based on Fermi-surface
integrals,\cite{wang-prb07} are in fact correct.

CBB start from their Eq.~(4), an expression for the intrinsic
AHC in terms of an integral of the Berry curvature over the
occupied band manifold in the Brillouin zone (BZ).  Following
Ref.~\onlinecite{wang-prb07} they write this as
\beq
\sigma_{xy} = \frac{1}{2\pi}\int_{-\pi}^{\pi} dk_z\,
   \sigma^{\rm 2D}_{xy}(k_z)\,,
\label{eq:sxy}
\eeq
where $\sigma^{\rm 2D}_{xy}$ is the contribution arising from a slice
of the BZ at a given $k_z$.
They then point out that if $\sigma^{\rm 2D}_{xy}$ is evaluated
as a sum of Berry phases computed as integrals over Fermi loops on
the slice,
\beq
\sigma^{\rm 2D}_{xy}(k_z) = \frac{e^2}{2 \pi h} \sum_n
\oint d\kk \cdot {\bf A}_{n\kk}(k_z)\,,
\label{eq:loop}
\eeq
where $\bf A$ is the Berry potential and the sum is over bands
crossing the Fermi energy, then contributions from entirely
filled bands can be missed.  Particularly when isolated
band crossings (``Weyl points'') are present in the occupied manifold,
they argue that Eq.~(\ref{eq:sxy}) will then yield an incorrect
result.

This is true as far as it goes.  However, the Fermi-surface formulas
proposed in Ref.~\onlinecite{haldane-prl04} \textit{are not} those
of Eqs.~(\ref{eq:sxy}-\ref{eq:loop}) above.  Instead, the formula
proposed in Eq.~(20) of Ref.~\onlinecite{haldane-prl04} states
that the non-quantized part of the AHC can be written, upon recasting
the Hall conductivity as a vector, as
\beq
\bm{\sigma}=\frac{e^2}{(2\pi)^2 h}\sum_\alpha \int_{S_\alpha}
d^2k \, [\bm{\mathcal F}(\kk)\cdot\hat{\bf n}(\kk)]\,\kk \,.
\label{eq:haldane}
\eeq
This takes the form of a sum of Fermi-surface integrals of the
position $\kk$ on the Fermi surface weighted by the surface-normal
component of the Berry curvature $\bm{\mathcal F}=\bm{\nabla}\times
{\bf A}$ of the band crossing the Fermi energy at $\kk$.  (The above
assumes that the Fermi sheets $S_\alpha$ do not touch the BZ boundary; the
generalization to the case that they do is provided in Eq.~(21) of
Ref.~\onlinecite{haldane-prl04}.) CBB seem to have overlooked that
this was the actual Fermi-surface expression proposed in
Ref.~\onlinecite{haldane-prl04}.  The possible existence of Weyl
points was carefully considered as part of the derivation of
Eq.~(\ref{eq:haldane}), which remains correct even when they are
present.

There is also no reason for concern that published first-principles
calculations of the AHC might be incorrect because of overlooking the
subtleties discussed by CBB.  Clearly those that were based on volume
integrals of the Berry curvature\cite{fang-science03,yao-prl04,wang-prb06}
are unaffected.  (In this class, approaches based on gauge-invariant
trace formulas\cite{lopez-prb12} are particularly suited to
the presence of Weyl points, since they remove the singularity
entirely.)

Of more concern is the Fermi-loop calculation of
Ref.~\onlinecite{wang-prb07}, which was also based on
Eqs.~(\ref{eq:sxy}-\ref{eq:loop}) above.  Since Berry phases are
only defined modulo $2\pi$, those equations must
be supplemented by a prescription for choosing the branch cuts
as a function of $k_z$.  CBB adopted a prescription in which the
sum of Berry phases in Eq.~(\ref{eq:loop}) was equated
with the 2D integral of the Berry curvature over the occupied
portions of the BZ for the partially filled bands only.  This leads
to unphysical step discontinuities in $\sigma^{\rm 2D}_{xy}$
at isolated $k_z$ values where a Weyl point between the last
fully occupied and the first partially occupied band crosses
the BZ slice, which CBB compensate for by adding a counterterm
in their Eq.~(8). Instead, in Ref.~\onlinecite{wang-prb07}
the quantity $\sigma^{\rm 2D}_{xy}$ was chosen to be
a continuous function of $k_z$.  In this way the extra
non-quantized contributions from filled bands in Eq.~(8) of CBB
are automatically included, as illustrated below.  In any case,
the results of the Fermi-loop and Fermi-sea integration approaches
were compared in Ref.~\onlinecite{wang-prb07} and found to agree.

As an instructive example, consider a nearly-insulating crystal that
is only metallic due to the presence of two small electron pockets
arising from shallow Weyl points of opposite chirality located at
$\kk_1$ and $\kk_2$.\cite{yang-prb11} 
In this case, Eq.~(8) of CBB includes a contribution to $\sigma_{xy}$
that is proportional to $(k_{2z}-k_{1z})$, which would be missed in a
naive implementation of
Eqs.~(\ref{eq:sxy}-\ref{eq:loop}) above.  This contribution is also
included in Eq.~(\ref{eq:haldane}), because the integral
of $\bm{\mathcal F}\cdot\hat{\bf n}$ over each Fermi
surface pocket is $\pm2\pi$, due to the enclosed Weyl
points; in the limit of small pockets, the factor of $\kk$
can be pulled out of the integral, providing
the needed $(k_{2z}-k_{1z})$ term.  In the
Fermi-loop approach of Ref.~\onlinecite{wang-prb07}, one sets the
branch choice of $\sigma^{\rm 2D}_{xy}$ arbitrarily at some
reference $k_z$, and then insists on continuity as a function of
$k_z$.  In this example, one can set $\sigma^{\rm 2D}_{xy}$ to
zero for $k_z$ below both pockets; it will then rise
continuously from 0 to $e^2/h$ while traversing the pocket around
$\kk_1$, then remain constant at $e^2/h$ until the second pocket is
reached, where it will again return to zero.  When averaged
over all $k_z$, this will correctly give a contribution
proportional to $(k_{2z}-k_{1z})$.

In summary, we conclude that the non-quantized part of the intrinsic
AHC is indeed correctly expressed as a Fermi-surface property in
Eqs.~(20-21) of Ref.~\onlinecite{haldane-prl04}, and that
the methods used in previous calculations of the AHC are correct,
even when Weyl points are present in the occupied band manifold.

%%=========================================================================


\end{document}